\documentclass[twocolumn]{aastex631}

\usepackage{amsmath}

\newcommand{\F}{\mathcal{F}}
\newcommand{\asini}{a_{\mathrm{p}}}

\def\check#1{{{\color{black}#1}}}
\def\tinytt#1{\tiny{\textrm{#1}}}

\newcommand\Tstrut{\rule{0pt}{2.9ex}}       % "top" strut
\newcommand\Bstrut{\rule[-1.3ex]{0pt}{0pt}} % "bottom" strut
\newcommand\TBstrut{\Tstrut\Bstrut}  

\usepackage{soul}

\graphicspath{ {./Plots/} {./} }
\begin{document}

\title{Constraints on r-modes and mountains on millisecond neutron stars in binary systems}

%\author{P.~B.~Covas$^{1}$, M. A. Papa$^{1}$, R. Prix$^{1}$}\noaffiliation
%\affiliation{Max Planck Institute for Gravitational Physics (Albert Einstein Institute), D-30167 Hannover, Germany}
\author[0000-0002-1845-9309]{P. B. Covas}
\affiliation{Max Planck Institute for Gravitational Physics (Albert Einstein Institute) and Leibniz Universit\"at Hannover\\
D-30167 Hannover, Germany}
\email{pep.covas.vidal@aei.mpg.de}
\author[0000-0002-1007-5298]{M. A. Papa}
\affiliation{Max Planck Institute for Gravitational Physics (Albert Einstein Institute) and Leibniz Universit\"at Hannover\\
D-30167 Hannover, Germany}
\email{maria.alessandra.papa@aei.mpg.de}
\author[0000-0002-3789-6424]{R. Prix}
\affiliation{Max Planck Institute for Gravitational Physics (Albert Einstein Institute) and Leibniz Universit\"at Hannover\\
D-30167 Hannover, Germany}
\email{reinhard.prix@aei.mpg.de}
\author[0000-0003-3919-0780]{B. J. Owen}
\affiliation{Department of Physics and Astronomy, Texas Tech University\\
Lubbock, Texas, 79409-1051, USA}
\email{benjamin.b.owen@ttu.edu}

%\input{git_tag.tex}
%\date{\commitDATE; \commitIDshort-\commitSTATUS}

\begin{abstract}
%\mapcomment{needs to be modified, after we've written the conclusion}
Continuous gravitational waves are nearly monochromatic signals emitted by asymmetries in rotating neutron stars. These signals have not yet been detected. Deep all-sky searches for continuous gravitational waves from isolated neutron stars require significant computational expense. Deep searches for neutron stars in binary systems are even more expensive, but potentially these targets are more promising emitters, especially in the hundreds-Hz region, where ground-based gravitational wave detectors are most sensitive.
% but are potentially more promising continuous wave emitters
%and get more so at higher frequencies, resulting in reduced sensitivity.
We present here an all-sky search for continuous signals with frequency between 300 and 500 Hz, from neutron stars in binary systems with orbital period between 15 and 60 days, and projected semi-major axis between 10 and 40 light-seconds. This is the only binary search on Advanced-LIGO data that probes this frequency range. Compared to previous results, our search is over an order of magnitude more sensitive. We do not detect any signals, but our results exclude plausible and unexplored neutron star configurations, for example, neutron stars with relative deformations greater than $3\times 10^{-6}$ within 1\,kpc from Earth and r-mode emission at the level of $\alpha \sim$ a few $10^{-4}$ within the same distance.
\end{abstract}

%% The AAS Journals now uses Unified Astronomy Thesaurus concepts:
%% https://astrothesaurus.org

%\maketitle
%\tableofcontents

\section{Introduction}

Detecting continuous gravitational waves is one of the most anticipated milestones in gravitational-wave astronomy. In spite of much effort, no detection has been achieved yet \citep{LIGOScientific:2021hvc,O3rmodesLIGO,O3SNRLIGO,O3allskyLIGO,O3accretingLIGO,O3aallskybinaryLIGO,Ashok:2021fnj,Zhang:2020rph,O2directedEaH,O2allskyFalcon,Steltner:2020hfd,Rajbhandari:2021pgc,Fesik:2020tvn}.

Continuous gravitational wave signals can be produced by asymmetric rotating neutron stars due to 
\begin{enumerate}
  \item[(i)] a \emph{mountain}, i.e. a non-axisymmetric deformation rigidly rotating with the star. If the principal moment of inertia is aligned with the spin axis of the star, this generates gravitational waves at frequency $f=2\,\nu$, where $\nu$ is the
    rotational frequency of the neutron star. %This maps our frequency search range to a neutron-star spin range of $\nu \in[150, 250]$\,Hz for the mountain model.

  \item[(ii)] \emph{r-modes}, a long-lasting oscillation mode that generates gravitational waves at a frequency of approximately $f\sim 4\,\nu/3$, with the exact relationship depending on the details of the neutron-star equation of state \citep{yoshida_r-mode_2005,idrisy_r_2015}.
%for which generally  are considered the most promising.
%    The r-mode emission model     The neutron star spin-range for r-mode emission covered by our search is therefore roughly
%    $\nu \in [225, 375]$\,Hz.
\end{enumerate}
Other mechanisms also exist, most notably following a glitch in the neutron star spin \citep[see][and references therein]{vanEysden:2008pd}, but we will not consider them here.

  The \emph{mountain} asymmetry could be generated by strains in the crust of the star, by internal magnetic
  fields or by accretion from a companion star \citep[see for example][]{asymmetry1,asymmetry2}.  Most theoretical studies in the literature \citep[e.g.][]{Owen_2005} provide \emph{upper limits} on the
  possible mountain height, while there are few concrete models \citep{Singh:2019dgy} predicting physical mechanisms
  that would actually produce such long-lasting mountains.

  On the other hand, \emph{r-modes} in rotating neutron stars are \emph{unstable} to gravitational-wave emission
  \citep{andersson_new_1998,friedman_axial_1998,Owen:1998xg,Lindblom:1998wf}. The instability is counteracted only by dissipative mechanisms in the neutron star such as viscosity, crust-core boundary friction, or through non-linear mode coupling \citep[e.g. see][]{bondarescu_spin_2007,bondarescu_spinning_2009,bondarescu_nonlinear_2013}. This leads to predictions of the so-called (in)stability window, i.e. ranges of the neutron star parameters where the r-modes are (un)stable.
  %and the growth of unstable r-modes is generally believed to be limited by non-linear mode coupling (e.g.\ see \cite{bondarescu_spin_2007,bondarescu_spinning_2009,bondarescu_nonlinear_2013}).  
 There are large
  theoretical uncertainties on the details of the r-mode stability window, amplitude and timescales, but
  several studies predict that long-lasting r-modes are present in accreting or post-accretion (quiescent)
  neutron stars in low-mass X-ray binaries (qLMXBs) \citep{Andersson:1998qs,Reisenegger:2003cq}.

  Furthermore, some authors \citep[e.g.][]{Chugunov:2014cwu,Maccarone:2022bym} argue for the existence of a potentially
  large unobserved population of quiescent LMXBs that may be subject to long-lasting ($\sim10^9$\,years)
  r-mode emission.
  These models provide compelling astrophysical motivation for an all-sky search for unknown neutron stars in
  binary systems emitting continuous gravitational waves. In turn, non-observation of such signals can
  potentially provide astrophysically relevant and informative upper limits, especially on r-mode amplitudes.

Independently of the emission mechanism, the expected continuous-wave amplitude is many orders of magnitude lower than the average noise level, so that months or years of data have to be combined together in order to build up a detectable signal-to-noise ratio.

The frequency of these signals is usually described in the source frame by an $n=1$ or $n=2$ order Taylor expansion around a reference time \citep[referred to as IT$n$ signals in][]{O2allskyFalcon}. This means that in the source frame 2 or 3 parameters are needed to describe the frequency evolution of the signal. In the observer frame, amplitude and Doppler modulations need to be accounted for in order to accurately describe the signal frequency evolution, which means that now the sky position of the source needs to be specified (2 more parameters). Searching for signals from neutron stars with known parameters does not pose any difficulty, since this amounts to a single waveform, that in the observer frame is specified by the value of 4 or 5 parameters. But for unknown sources, the high resolution in all the parameters makes the all-sky, broad frequency searches the most computationally costly among all gravitational-wave searches. A recent review of different methods to perform these searches can be found in \citet{Tenorio:2021wmz}.

% Astrophysical motivation for all-sky binary searches, in particular for higher frequency ones.
If the unknown neutron star is assumed to be in a binary system rather than being isolated, the search problem becomes even more difficult. Assuming a circular orbit, three additional parameters need to be accounted for, which further increases the already high computational cost of all-sky surveys. However, since more than half of the known millisecond pulsars are in binary systems, and their accretion history offers channels to generate the asymmetries needed for a detectable continuous signal \citep{Holgado_Ricker_Huerta_2018, Gittins_Andersson_2019, Singh_Haskell_Mukherjee_Bulik_2019,Chugunov:2014cwu}, searches for signals from unknown neutron stars in binaries are very interesting.

In this paper we present the results of an all-sky search for continuous waves from neutron stars in binary systems with gravitational-wave frequencies between 300 and 500 Hz, using the public Advanced LIGO O3a data \citep{GWOSCO3a}. This search complements two previous Advanced LIGO data searches which looked at frequencies between 50 and 300 Hz \citep{O2aallskybinary, O3aallskybinaryLIGO}. The frequency range up to 500\,Hz was previously explored using S6 and VSR2/3 data \citep{TwoSpectS6}, producing upper limits on the gravitational-wave amplitude not more stringent than $\sim 7.5 \times 10^{-24}$. Our search improves on these results by nearly a factor of 30, with our most restrictive upper limit being $\sim 2.8 \times 10^{-25}$.

We search for signals from systems with projected semi-major axis between 10 and 40 light-seconds, and orbital periods between 15 and 60 days. \check{These parameter ranges are very similar to the ones searched by \cite{O2aallskybinary} and in region A of \cite{O3aallskybinaryLIGO}, the difference being that we have extended the largest orbital period from 45 to 60 days. The previous search using S6 and VSR2/3 data \citep{TwoSpectS6} explored a wider range of parameter space, covering orbital periods from 0.08 to 94 days.}

We use a newly-developed search pipeline -- \texttt{BinarySkyHou$\mathcal{F}$} -- which combines coherent $\F$-statistic search results with the Hough Transform method \citep{krishnan04:_hough,2019PhRvD..99l4019C}. \texttt{BinarySkyHou$\mathcal{F}$} is an improvement over the method of \citet{2019PhRvD..99l4019C}, due to its lower computational cost and the usage of a \check{generally} more sensitive coherent detection statistic \citep{BSHFstat}.

The paper is organised as follows: in Section \ref{sec:data} we describe the data that we use; in Section \ref{sec:method} we briefly explain the search method; in Section \ref{sec:results} we present results and discuss their astrophysical implications, which we summarize in Section \ref{sec:conclusions}.

\section{Data}
\label{sec:data}

\begin{figure}
  \begin{center}
    \includegraphics[width=1.0\columnwidth]{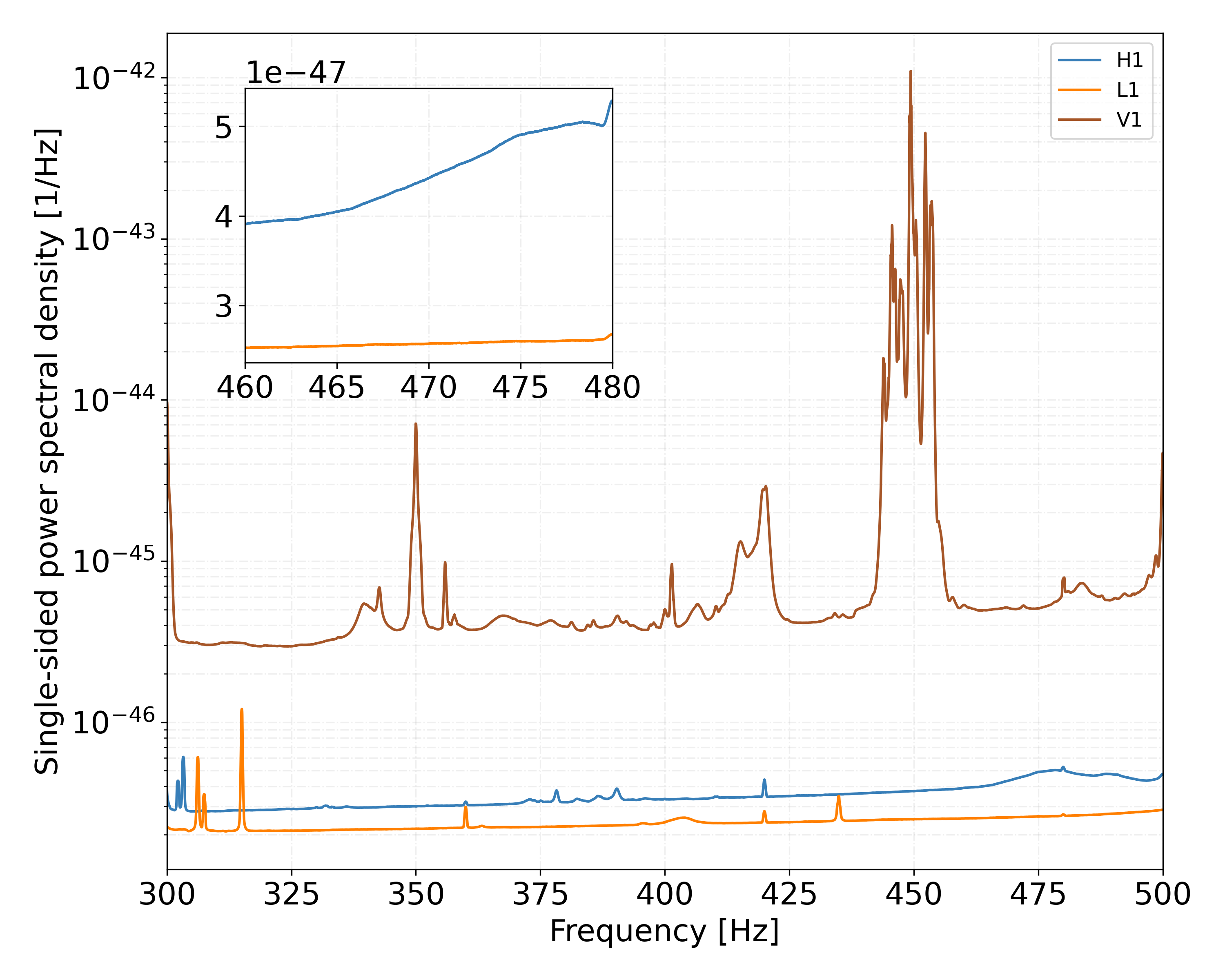}
    \caption{Power spectral density $S_{\mathrm{n}}$, harmonically averaged over SFTs, for the three detectors as a function of frequency. The inset shows the region around the surviving outlier of the follow-up.}
    \label{fig:PSD}
  \end{center}
\end{figure}

This search uses the Advanced LIGO O3a data \citep{GWOSCO3a}. O3a comprises the first 6 months of the O3 run, i.e. April-October 2019, and is the first O3 dataset publicly released. We use data from both LIGO gravitational-wave detectors, i.e. from the Hanford (H1) and the Livingston (L1) detectors \citep{LIGOScientific:2014pky}. We do not consider data from the Virgo detector because the amplitude spectral density is $\sim 3$ times higher than that of LIGO and is more contaminated by non-Gaussian disturbances. The harmonic average power spectral density of the LIGO and Virgo detectors in the frequency range covered by this search is shown in Figure~\ref{fig:PSD}. 

We use the \texttt{GWOSC-16KHZ\_R1\_STRAIN} channel and the \texttt{DCS-CALIB\_STRAIN\_CLEAN\_SUB60HZ\_C01\_AR} frame type. This data \check{had} been cleaned \check{before public release} at the frequencies of the power 60\,Hz lines and the calibration lines. Many other lines are still present in the dataset, which produce spurious outliers that need to be examined and discarded, as will be seen in Section \ref{sec:results}. %We use the publicly available list of lines collected by the LVC \todo{CITE}.

% Gating
As discussed in other publications \citep[e.g.][]{O3allskyLIGO}, the O3a dataset suffers from a large number of short-duration glitches that increase the noise level in the frequency range of interest for this search, thus potentially reducing the sensitivity. For this reason, we use the gating method developped by \citet{gating} to remove these glitches and improve the data quality.

The dataset of each detector is divided in short Fourier transforms
\citep[SFTs, see][]{Allen_Mendell} of 200 s, \check{which is short enough that the signal power does not spread by more than a 1/200 Hz frequency bin during this time. In fact, in the chosen range of orbital and frequency parameters, this remains the case for durations up to $\approx $ 1200 s \citep[see Figure 8 of][]{2019PhRvD..99l4019C}}. The total number of SFTs is 55068 for H1 and 56118 for L1. 

\section{The search}
\label{sec:method}

\subsection{Signal model}
The gravitational-wave signal $h(t)$ from a tri-axial asymmetric rotating neutron star as a function of time $t$ in the observer frame is given by \cite{jaranowski_data_1998}:
\begin{eqnarray}
  h(t) = h_0 & \left[ F_+(t; \psi) \frac{1 + \cos^2{\iota}}{2} \cos{[\phi_0 + \phi(t)]} \right. \nonumber \\
  + & \left. F_{\times}(t; \psi)  \cos{\iota} \sin{[\phi_0 + \phi(t)]} \right]
\end{eqnarray}
where $F_+$ and $F_{\times}$ are the antenna-pattern functions of the detector, $\iota$ is in the inclination angle of the angular momentum of the source with respect to the line of sight, $\psi$ is the polarization angle, $\phi_0$ is the initial phase, $\phi(t)$ is the gravitational-wave phase at time $t$ and $h_0$ is the intrinsic gravitational wave amplitude. %of the signal is:

The phase of the signal $\phi(t)$ in the observer frame depends on the intrinsic frequency $f_0$ and frequency derivatives of the signal and on the Doppler modulation of this signal due to the relative motion between the source and the detector. The Doppler modulation depends on the gravitational-wave frequency, the sky position of the source ($\alpha$ and $\delta$), and the parameters describing the Keplerian orbit of the neutron star. These orbital parameters are: orbital period $P$, projected orbital semi-major axis $\asini$, time of ascension $t_{\mathrm{asc}}$, argument of periapsis, and eccentricity. In an all-sky search for IT$1$ signals this in principle amounts to $9$ parameters to be searched for, in order to accurately track the signal. 

\subsection{Parameter space covered}

We search for signals with intrinsic frequency $f_0$ between 300 and 500\,Hz and frequency derivative $|\dot{f}_0| \le 4 \times 10^{-10}$\,Hz/s, across the entire sky. We choose this frequency range because it has not yet been covered by a search on Advanced-LIGO data. The chosen frequency derivative range allows to \check{set all template waveform $\dot{f}_0 =0$ and hence not search over multiple different frequency derivative values in the first stage of the search. With this range we are able to cover the expected frequency derivative values produced by spin-wandering, which may appear in accreting systems \citep{spinwandering}}.

\begin{figure}
  \begin{center}
    \includegraphics[width=1.0\columnwidth]{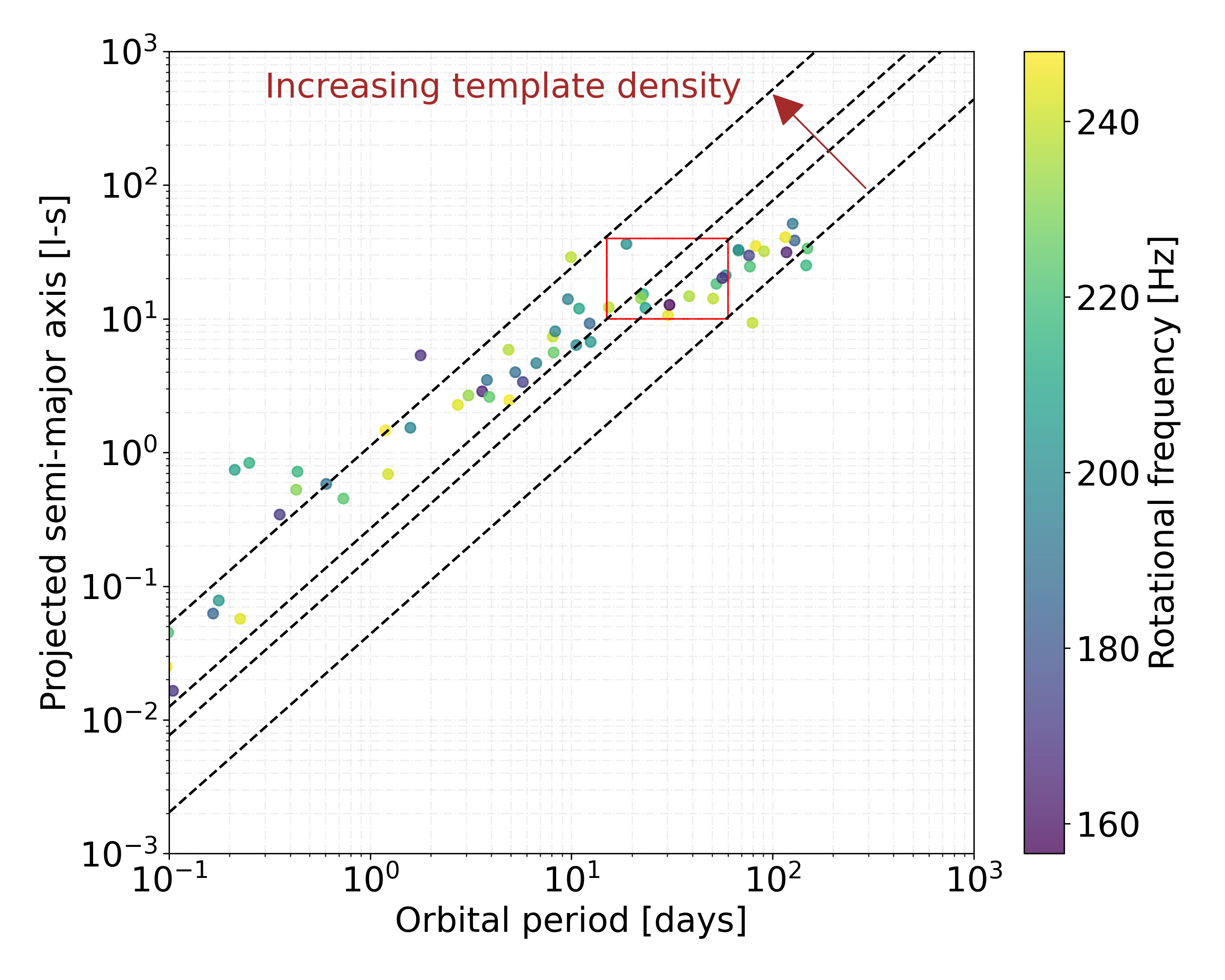}
    \caption{The red box shows the parameter space covered by this search, while the points show the population of known pulsars in binary systems with rotational frequencies between 150 Hz and 250 Hz. \check{The dashed black lines indicate points of equal template density.}}
    \label{fig:ParameterSpace}
  \end{center}
\end{figure}

We assume that the signals come from a neutron star in a binary system. As shown in Figure~\ref{fig:ParameterSpace}, neutron stars in binary systems cover a broad range of orbital separations $\asini$, depending on the neutron star companion mass and on the orbital period. On the other hand, all-sky high-sensitivity/high-resolution searches over such a broad range of orbital parameters are impossible due to their computational cost, so we concentrate the search to the range of $P-\asini$ indicated by the red box in Figure~\ref{fig:ParameterSpace}. This box is chosen in such a way that, among the search-boxes that could be drawn in that plane with the same computational cost for the associated search, this box contains the highest number of pulsars in it, \check{while probing higher orbital periods than previous searches on Advanced LIGO data \cite{O3aallskybinaryLIGO}. To give some idea of how the computational cost varies in the $\asini-P$ plane we also plot lines of constant template-count density \citep[from Eq.~77 of][]{Leaci:2015bka}.}
%this is the box with the highest number of known neutron stars in it. 
Following \citet{2019PhRvD..99l4019C} we assume an orbital eccentricity smaller than $5.7 \times 10^{-3}~[{500~\textrm{Hz}\over {f_0}}]$, so in the first stage of the search we  \check{set all template waveform $e =0$ and not search over different eccentricity (and argument of periapsis) values}.

The \check{signal parameter space covered by the search} is summarized in Table \ref{tab:region}. The total number of templates searched is $\sim 6 \times 10^{14}$.
\begin{deluxetable}{lccc}
\tablecaption{
This table shows the \check{the range of values of the signal waveform parameters covered by the search. As explained in the text, in order to cover the first frequency derivative ${\dot{f_0}}$ and the eccentricity $e$ ranges, we only search a single template. For this reason we do not quote a grid resolution for these two parameters in  Table \ref{tab:setup}.}  $t_m$ is the mid-time of the search.
\label{tab:region}
}
\tablehead{
\colhead{Parameter} & \colhead{Range}
}
\startdata
%\TBstrut $T_{\mathrm{seg}}$ [s] & $900$ & $900$ \\ 
\TBstrut $f_0$ : Frequency [Hz] & 300 - 500 \\ 
\TBstrut $|{\dot{f_0}}|$ : Frequency deriv. [Hz/s] & $< 4 \times 10^{-10}$ \\ 
\TBstrut $a_{\mathrm{p}}$ : Projected semi-major axis [l-s] & 10 - 40 \\ 
\TBstrut $P$ : Orbital period [days] & 15 - 60 \\ 
\TBstrut $t_{\mathrm{asc}}$ : Time of ascension [s] & $t_m \pm P/2$ \\ 
\TBstrut $e$ : Orbital eccentricity & $ < 5.7 \times 10^{-3}~[{500~\textrm{Hz}\over {f_0}}]$ \\ 
\TBstrut $\alpha$ : Right ascension [rad] & 0 - $2\pi$ \\ 
\TBstrut $\delta$ : Declination [rad] & $-\pi/2$ - $\pi/2$ \\ 
\enddata
\end{deluxetable}
%%%%%%
%For this reason, if the neutron star emitting gravitational waves is accreting during our search, there is an accretion rate (maybe combined with other parameters) to which we are not sensitive too, since it would cause a $f_1$ value higher than what we are sensitive to \todo{Calculate that value}.

\subsection{Summary of pipeline}

We use a semi-coherent search method, where the data is separated in segments of a maximum span $T_{\mathrm{seg}} =900$\,s. For all segments we calculate the coherent detection statistic $\mathcal{F}$ of \cite{Cutler:2005hc} and \cite{jaranowski_data_1998} over a coarse grid in frequency and sky. The segment length $T_{\mathrm{seg}}$ is short enough that the orbital parameters are not resolved. The $\mathcal{F}$-statistic values are then combined with an improved version of the \texttt{BinarySkyHough} methodology of \cite{2019PhRvD..99l4019C}, described in more detail in \citet{BSHFstat}. \check{By using the $\mathcal{F}$-statistic, the new pipeline can leverage sensitivity gains from the use of  longer coherent duration segments and from the coherent combination of data from different detectors. However, with the particular set-up used for this search -- short segment length and two detectors -- its sensitivity is comparable to that of power-based statistics.}

The final detection statistic is a weighted sum of the single-segment statistic $\F_i$: 
\begin{equation}
\label{eq:twoFsum}
  2\mathcal{F}_{\mathrm{sum}} = \sum_{i=1}^{N_{\mathrm{seg}}} w_i \,2\mathcal{F}_i,
\end{equation}
where $w_i$ depends on the detectors' antenna beam-pattern functions during segment $i$, inverse-weighted with the noise power spectral density, and $N_{\mathrm{seg}}=16\,966$. \check{The sum of Eq.~\ref{eq:twoFsum} depends on the template orbital parameters.}
%, and 
%\begin{equation}
%  w_i \propto \frac{A_i+B_i}{S_{\mathrm{n};i}},
%\end{equation}
%where $A$ and $B$ are the antenna-pattern matrix coefficients (calculated for the given sky point in the coherent grid) \citep[see][]{jaranowski_data_1998}. 

We use the coherent detection statistic $\F$ for all segments that have a well-conditioned antenna pattern matrix with condition number $<10^4$. For larger condition numbers the $\F$-statistic becomes numerically singular and we use the constant antenna-pattern detection statistic with two degrees of freedom derived in \cite{2DOF}.

\begin{figure}
  \begin{center}
    \includegraphics[width=1.0\columnwidth]{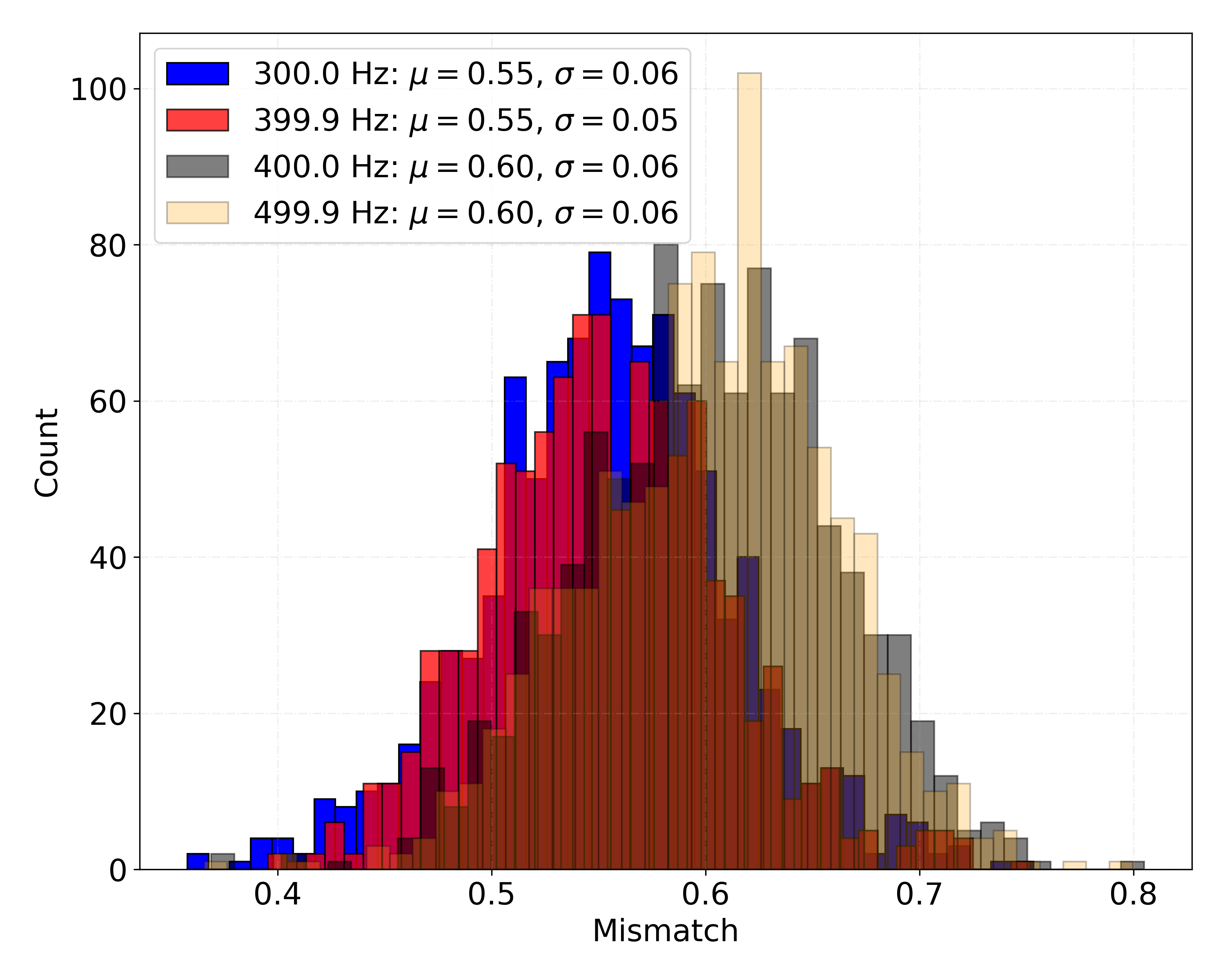}
    \caption{The mismatch distribution of this search at four
      frequencies, \check{with the other parameters spanning the ranges given in Table \ref{tab:region}, as explained in the text}.}
    \label{fig:Mismatch}
  \end{center}
\end{figure}

\begin{deluxetable}{lccc}
\tablecaption{
\check{Grid} resolutions in the various waveform parameters. $\Omega=2\pi/P$ is the average angular orbital velocity. \label{tab:setup}
}
\tablehead{
 & & \multicolumn{2}{c} {Frequency Range}\\
Resolution &  & \colhead{[300,400)} & \colhead{[400,500)}
}
\startdata
%\TBstrut $T_{\mathrm{seg}}$ [s] & $900$ & $900$ \\ 
\TBstrut $\delta f_0$ [mHz] & & $1.1$ &  $1.1$\\ 
%\TBstrut \check{$|\delta{\dot{f_0}}|$} \check{[Hz/s]} & & \check{$4 \times 10^{-10}$}  & \check{$4 \times 10^{-10}$} \\
\TBstrut $\delta \asini $ [l-s] $[{{P \over {37 \tinytt{days}}}}]$ & & $ 2.7 ~[{{400 ~ \tinytt{Hz}}\over{f_0}}] $ &  $3.1~[{{500 ~ \tinytt{Hz}}\over{f_0}}] $\\
\TBstrut $\delta \Omega ~[10^{-8}$ rad] $[{{P \over {37 \tinytt{days}}}}] ~ [{{25 ~ \tinytt{l-s}}\over{\asini}}] $ & & $ 2.4 ~[{{400 ~ \tinytt{Hz}}\over{f_0}}] $ &   $2.7~[{{500 ~ \tinytt{Hz}}\over{f_0}}] $\\
\TBstrut $\delta t_{\mathrm{asc}} ~[10^{4}$ s] $[{{P \over {37 \tinytt{days}}}}]^2 ~ [{{25 ~ \tinytt{l-s}}\over{\asini}}] $ & & $ 5.5 ~[{{400 ~ \tinytt{Hz}}\over{f_0}}] $ &   $6.2~[{{500 ~ \tinytt{Hz}}\over{f_0}}] $\\
%\TBstrut \check{$\delta e ~ [10^{-3}]$} & & \check{$4.6 ~[{400~\textrm{Hz}\over {f_0}}]$} & \check{$  5.7 ~[{500~\textrm{Hz}\over {f_0}}]$} \\
\TBstrut $\delta \alpha =\delta \delta  ~[10^{-2}$ rad ]& & $4.3 ~[{{400 ~ \tinytt{Hz}}\over{f_0}}]$ &  $4.0~[{{500 ~ \tinytt{Hz}}\over{f_0}}]$\\
& & & \\
\enddata
\end{deluxetable}
We divide the search into a low- and a high-frequency region, each 100\,Hz wide, and use different template grids in order to balance the computational cost in the two regions. The grid resolutions for all the parameters are shown Table \ref{tab:setup}.
The overall resulting average mismatch is $\lesssim 0.6$ and the mismatch distributions at four frequencies spanning the search range are shown in Figure~\ref{fig:Mismatch}.

We define a significance $s$ for every search result as
\begin{equation}
  s = \frac{2\mathcal{F}_{\mathrm{sum}} - \mu}{\sigma},
\label{eq:sig}
\end{equation}
where
\begin{align}
\label{eq:mu}
  \mu &= 4 \sum\limits_{i=1}^{N_4} w_{4;i} + 2 \sum\limits_{j=1}^{N_2} w_{2;j}, \\
  \label{eq:sigma}
  \sigma^2 &= 2 \,\left( 4 \sum\limits_{i=1}^{N_4} w^2_{4;i} + 2 \sum\limits_{j=1}^{N_2} w^2_{2;j}\right),
\end{align}
are the mean and standard deviation of $2\mathcal{F}_{\mathrm{sum}}$ in Gaussian noise, and $w_{4;i}$ ($w_{2;j}$) are the weights for the $N_4 (N_2)$ segments with four (two) degrees of freedom, respectively.

\section{Results}
\label{sec:results}

\subsection{Outliers, follow-up}

\begin{figure}
  \begin{center}
    \includegraphics[width=1.0\columnwidth]{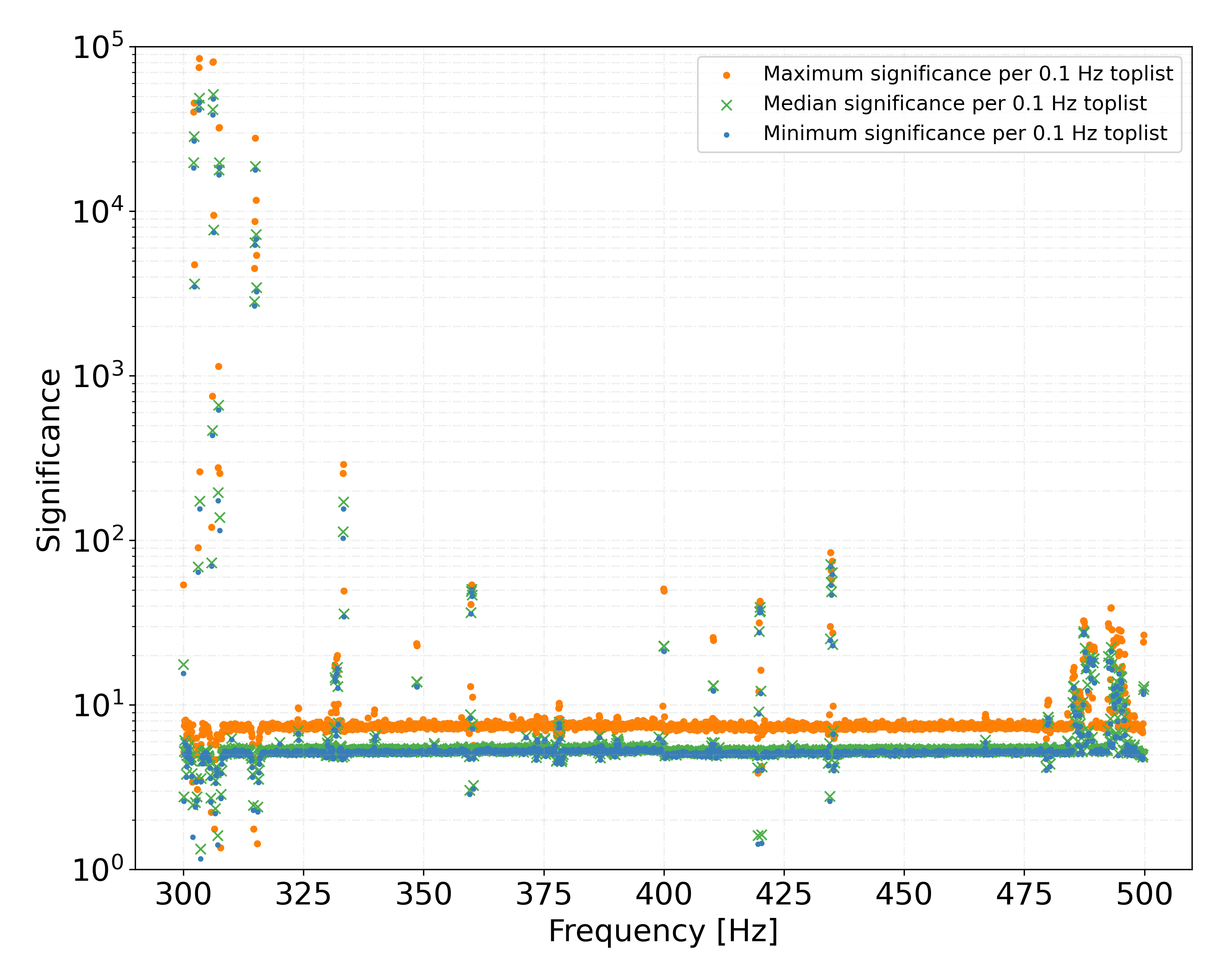}
    \caption{Significance of the top $5 \times 10^{5}$ candidates in each 0.1\,Hz frequency band. We show the maximum, median, and minimum values. The lower-than-average values of the significance are due to the $\sigma$ from  Eq.~\ref{eq:sigma} being an overestimate of the actual standard deviation, due to the noise weight being higher in the vicinity of a noise disturbance.}
    \label{fig:Toplists}
  \end{center}
\end{figure}

We consider the most significant $5 \times 10^{5}$ results in every 0.1\,Hz band, which are shown in Figure~\ref{fig:Toplists}. On these we use the clustering procedure of \citet{2019PhRvD..99l4019C,O2aallskybinary} and group together results due to the same cause. The top ten clusters per 0.1\,Hz band are saved for further analysis, yielding a total of 20\,000 clusters. In what follows we will refer to these selected clusters as {\it{candidates}}.

%Figure \ref{fig:Toplists} shows the significance results per each searched 0.1 Hz band. At each band we use the clustering algorithm explained in \citet{2019PhRvD..99l4019C,O2aallskybinary}, in order to identify clusters of candidates in the six-dimensional parameter space. 

Before analyzing the candidates with a longer coherence time, we search for the presence of instrumental lines in their vicinity. We use the list of LIGO detectors' lines compiled by \citet{lines}. We calculate the frequency-time pattern of every candidate and discard it if it crosses any of the lines in the lists. Doing this, 19\,044 candidates remain.

In the next step we apply the same follow-up that was used in \citet{O2aallskybinary}, where we increase the coherence time ($T_{\textrm{seg}}$) to 9000\,s and use a Markov Chain Monte Carlo (MCMC) procedure to calculate the detection statistic \citep{Keitel2021,pyfstat}. Due to the significantly longer coherent time baseline of the follow-up, the detection statistic $\overline{2\F}$ is an average of the detection statistics from each of the 1\,728 segments, without any noise weights. Furthermore, \check{due to the increase of coherence time, the follow-up resolves different values of the argument of periapsis, eccentricity, and frequency derivative, which now need to be all searched for explicitly}. %the initial search covers a range (discussed in section \ref{sec:method}) even when not explicitly searching, but  these parameters need to be explicitly included in the MCMC follow-up. %in order to cover the same ranges.

\begin{figure}
  \begin{center}
    \includegraphics[width=1\columnwidth]{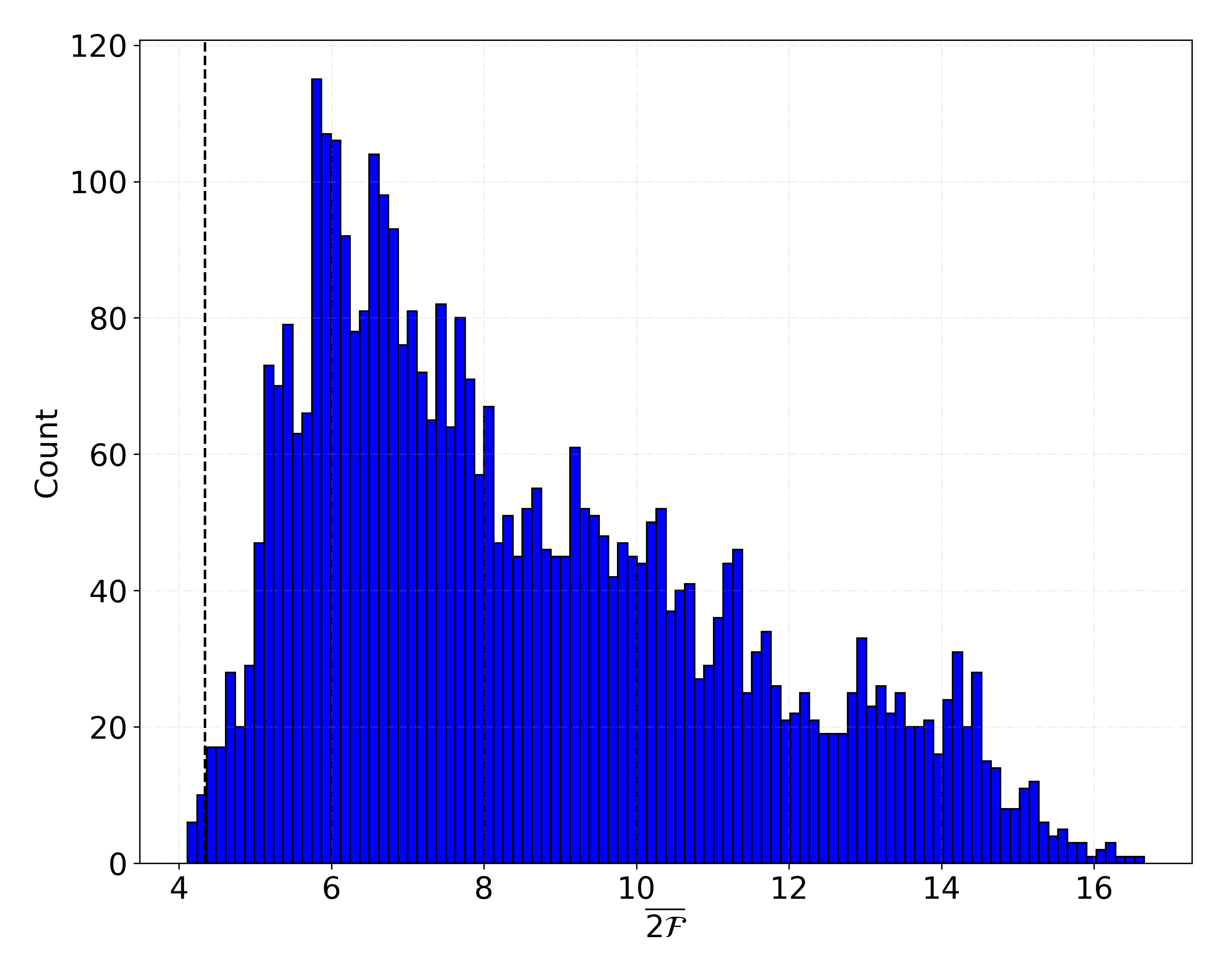}
    \caption{Distribution of $\overline{2\F}$ values (averaged over segments) from the MCMC follow-up of 4\,763 fake signals that survive the initial stage of our search.}
        \label{fig:FUInjections}
  \end{center}
\end{figure}

If one of our MCMC candidates is due to a gravitational-wave signal its detection statistic value must be consistent with what is expected from a signal. This is shown in Figure~\ref{fig:FUInjections}, where we have the distribution of detection statistic values after the MCMC follow-up of 4\,763 candidates stemming from fake signals added to the O3a data. 
\check{The signal parameters are distributed within the ranges of Table~\ref{tab:region} \footnote{\check{The cosine of the source inclination angle, the polarization angle, the initial phase, the frequency, the orbital period, the projected semi-major axis, the log of the eccentricity and of the frequency derivative are uniformly distributed; sky positions are isotropically distributed in the sky.}}}. 
Each of the 4\,763 candidates has passed the previous stage of the search, i.e. is one of the top 10 clusters in its 0.1 Hz band. The amplitudes of the fake signals bracket the 95\% upper limit value. Based on the distribution of recovered values we set a threshold for our MCMC candidates at $\overline{2\F}_{\textrm{thr}}=4.34$, which corresponds to about a 1\% false dismissal rate.  

\begin{figure}
  \begin{center}
     \includegraphics[width=1\columnwidth]{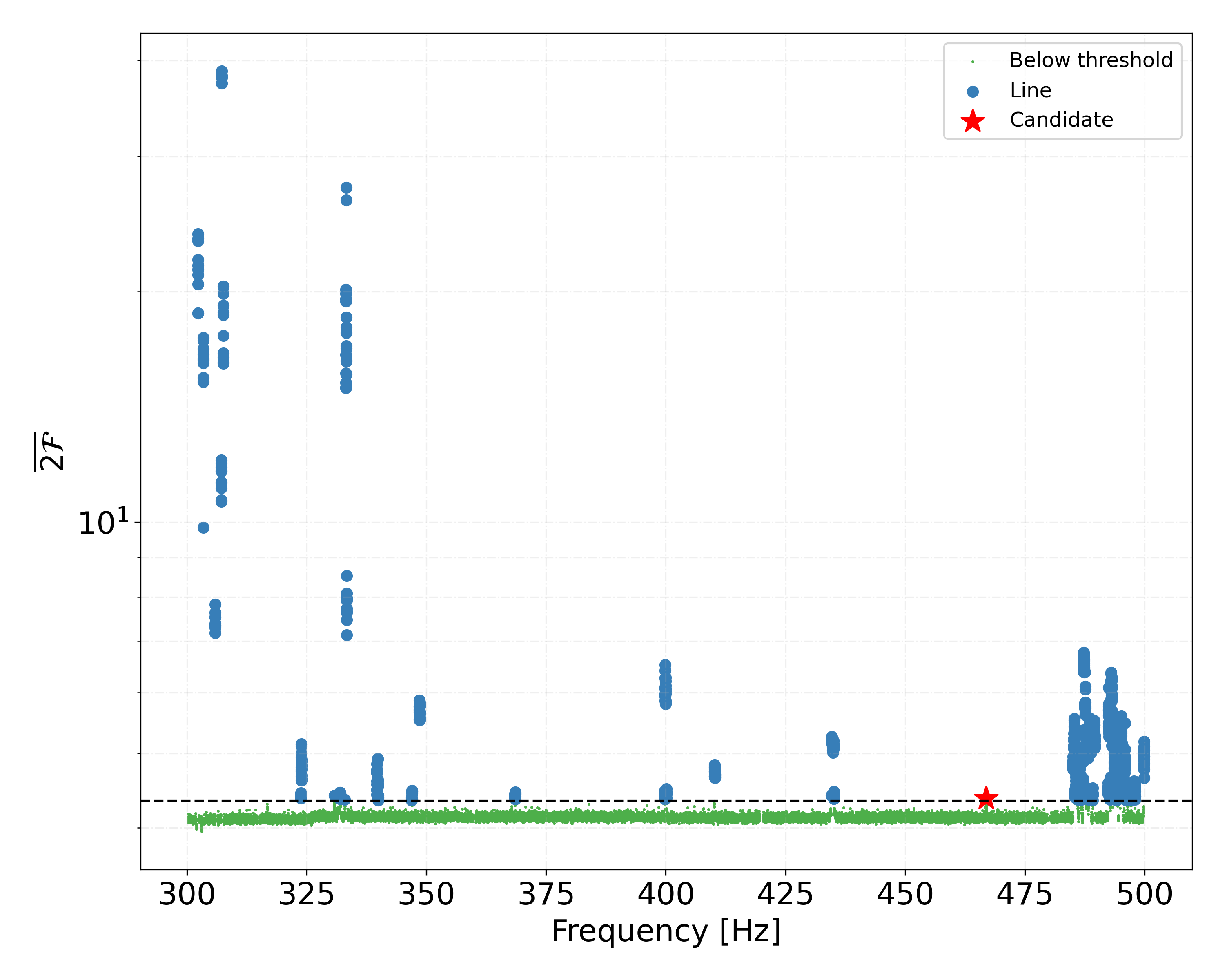}
    \caption{$\overline{2\F}$ values for each of the 19\,044 candidates followed-up with the MCMC procedure, as a function of the frequency of the candidate signals. The horizontal threshold line is the threshold value at $\overline{2\F}_{\textrm{thr}}=4.34$.}
        \label{fig:Followup}
  \end{center}
\end{figure}

\begin{figure}
  \begin{center}
    \includegraphics[width=1.0\columnwidth]{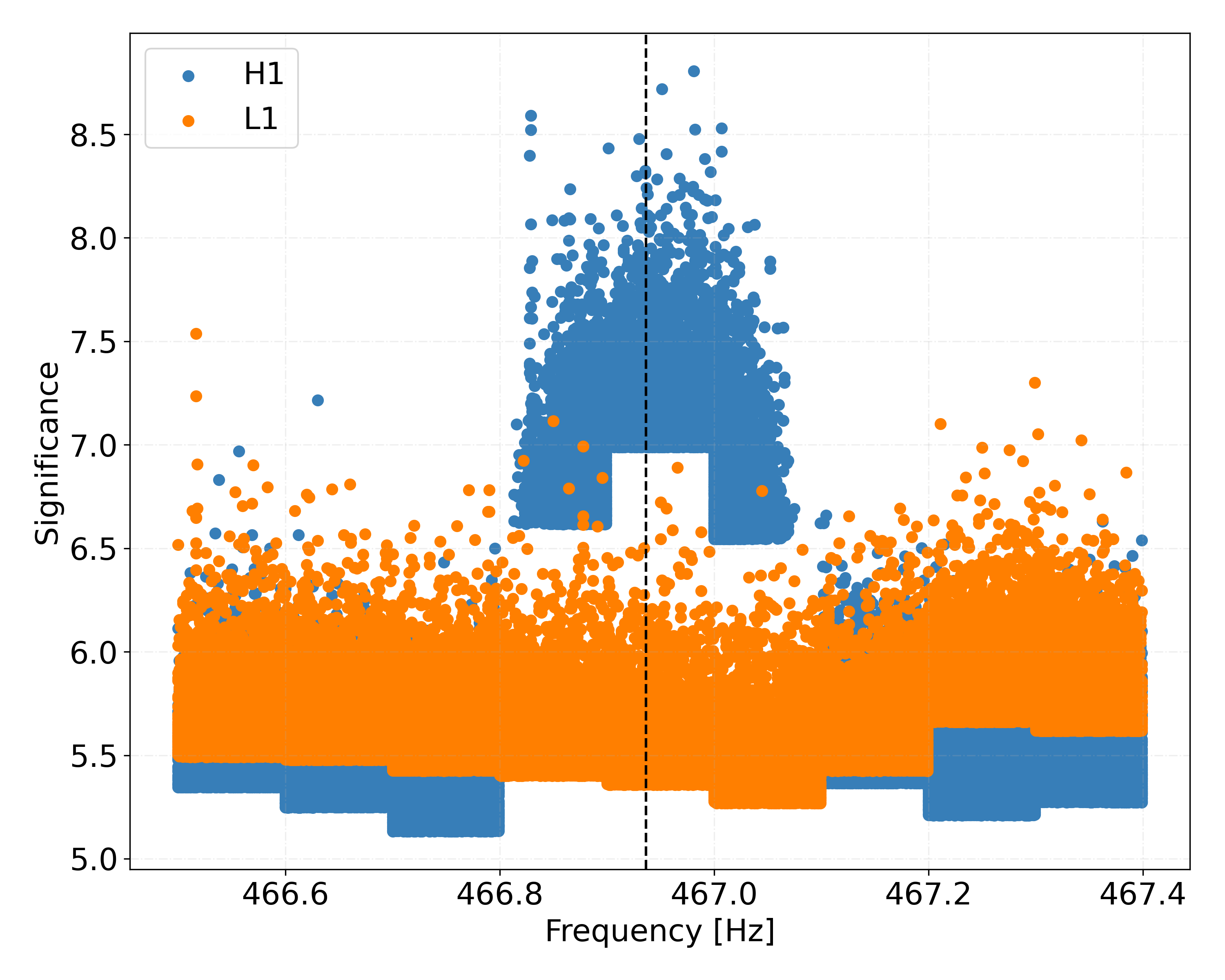}
    \caption{Significance values around the candidate at 466.93\,Hz when analyzing only H1 or L1 data. The vertical dashed line marks the frequency of the outlier that survives the follow-up stage.}
    \label{fig:Candidate}
  \end{center}
\end{figure}

The distribution of $\overline{2\F}$ for our candidates is shown in Figure~\ref{fig:Followup}: only one candidate out of the 19\,044 survives, at $f_0\sim 466.93$\,Hz. 
We investigate the candidate by performing single-detector searches with the same set-up as the initial search. This reveals higher detection statistic values from the H1 detector over a too broad range of frequencies ($\sim 0.3$\,Hz) to be consistent with an astrophysical signal (see Figure~\ref{fig:Candidate}). Even though this spectral region does not display significant disturbances (see inset of Figure~\ref{fig:PSD}) and no line is present in the line-list \citep{lines}, the fact that the high detection statistic value comes from the least sensitive detector, further supports the idea that it is not of astrophysical origin. Additionally, we perform a $T_{\textrm{seg}}=3600$\,s dedicated \texttt{BinarySkyHou$\mathcal{F}$} follow-up of this candidate with only L1 data (i.e.\ the undisturbed data), and the resulting detection statistic falls short of what is expected for a signal. Therefore no candidates survive the follow-up stage. 

We note that no hardware injections (artificially added signals) are present in the frequency range covered by this search, so we do not expect any surviving candidates from hardware-injected signals.

\subsection{Upper limits}
We calculate the 95\%-confidence upper limits on the intrinsic gravitational-wave amplitude $h_0^{95\%}(f_0)$ in every 0.1\,Hz band. This is the amplitude such that 95\% of a population of signals with frequency in that band and with the values of the other parameters in our search range,  would have been detected. The detection criterion for a signal is that \check{it makes it in the top-ten list for the relevant band and that it passes the MCMC-output threshold.}

We determine the $h_0^{95\%}(f_0)$ in four representative 0.1\,Hz bands in the lower and upper frequency regions. In each band we compute the percentage of detected signals from 500 search-and-recoveries, with signals added to the data at a fixed value of $h_0$. %We use values of $h_0$ that bracket the 96\% detection efficiency\footnote{We need 96\% detection efficiency due to the 1\% false dismissal from the $\overline{2\F}_{\textrm{thr}}$ threshold.}, which is the point at which we measure the $h_0^{95\%}$ upper limit, with a sigmoid fit. 
This is a standard procedure, recently also employed by \cite{O3aallskybinaryLIGO,O2aallskybinary,Steltner:2020hfd}.

%At 8 different 0.1 Hz frequency bands we add 500 signals to the real data, and calculate the percentage of detected signals at three different sensitivity depth values, given by $\sqrt{S_n} / h_0$. We find the sensitivity depth where we detect 95\% of the signals with a sigmoidal fit, and from this sensitivity depth we can calculate the $h_0^{95\%}$ values. 
We can associate to each 0.1\,Hz upper limit a corresponding sensitivity depth ${\mathcal{D}}$ \citep{Behnke:2014tma,Dreissigacker:2018afk}, defined as
\begin{equation}
\label{eq:sensDeph}
{\mathcal{D}}^{95\%}(f)=\frac{\sqrt{S_{\mathrm{n}}(f)}}{h_0^{95\%}},
\end{equation}
where $S_{\mathrm{n}}(f)$ is the power spectral density of the data. In reasonably well-behaved noise this quantity does not vary much with frequency, and is a good measure of the sensitivity of the search since it quantifies how far below the noise floor the smallest detectable signal lies. The average sensitivity depth for the low- and high-frequency band of Table \ref{tab:setup} is $17.6^{+1.3}_{-1.4}~[1/\sqrt{\textrm{Hz}}]$ and $16.2^{+1.7}_{-1.9}~[1/\sqrt{\textrm{Hz}}]$, respectively, with the $1\sigma$ uncertainties.
We use these sensitivity depth values as scale factors to determine the upper limits in the remaining bands, from measurements of the noise floor $S_{\mathrm{n}}$. The upper limit values in machine-readable format are available in the Supplementary Materials\footnote{Upper limit values in machine-readable format are also available at \texttt{https://www.aei.mpg.de/continuouswaves/O3aAllSkyBinary}}.

\begin{figure}
  \begin{center}
    \includegraphics[width=1.0\columnwidth]{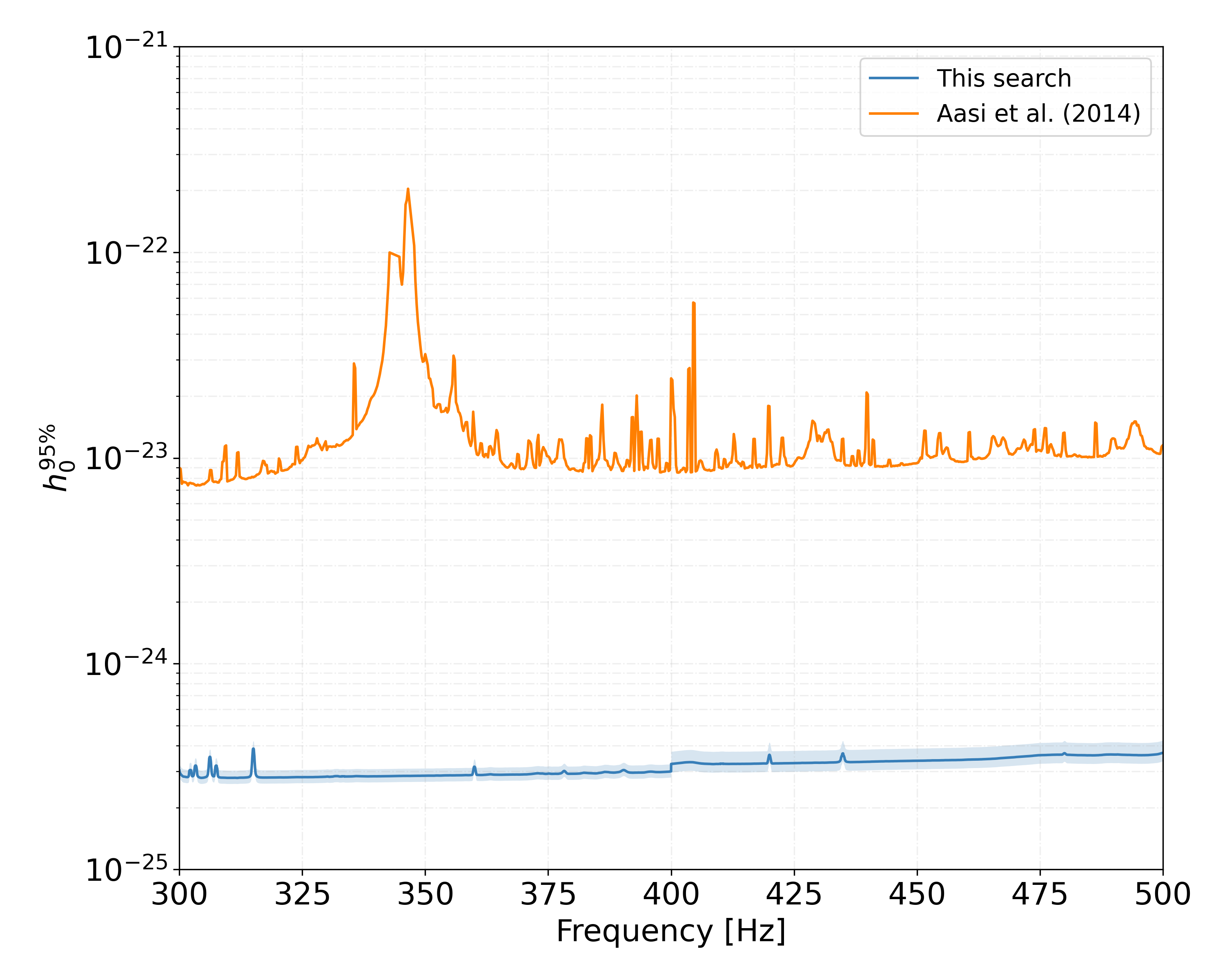}
    \caption{Upper limits on the gravitational-wave amplitude $h_0^{95\%}$. The lower blue curve shows the results for this search (with the shaded area showing the $1\sigma$ confidence region), while the upper orange curve shows the S6 data results of \cite{TwoSpectS6}, the only previous search covering this frequency range. \check{The slight discontinuity at $400$ Hz is due to the change in the search grid spacings, as shown in Table \ref{tab:setup}.}}
    \label{fig:UL}
  \end{center}
\end{figure}

%Comparison with LVC O3a results. Comparison with isolated all-sky searches and targeted.
The results of this procedure are shown in Figure~\ref{fig:UL}. The most sensitive upper limit is of $2.8 \times 10^{-25}$ at 311.7\,Hz. Compared with the only previous search that analyzed this region of parameter space using S6 data \citep{TwoSpectS6}, this search is over 27 times more sensitive, due to the more sensitive dataset and a more sensitive pipeline. \cite{O3aallskybinaryLIGO} also searched O3a data (but only up to 300\,Hz) and attained a sensitivity comparable to ours.%, although we employ a coarser grid resolution.

\subsection{Astrophysical reach}

The gravitational-wave amplitude emitted due to an equatorial ellipticity $\epsilon$ is
\begin{equation}
  h_{0} = \frac{4 \pi^{2} G}{c^{4}} \frac{I_{z z} \epsilon f_0^{2}}{d},
  \label{eq:h0}
\end{equation}
where $\epsilon=|I_{xx} - I_{yy}|/I_{zz}$, $I_{zz}$ is the moment of inertia of the star with respect to the principal axis aligned with the rotation axis, $d$ is the distance, and $f_0$ is the gravitational-wave frequency, which in this model is equal to twice the rotational frequency.

\begin{figure}
  \begin{center}
    \includegraphics[width=1.0\columnwidth]{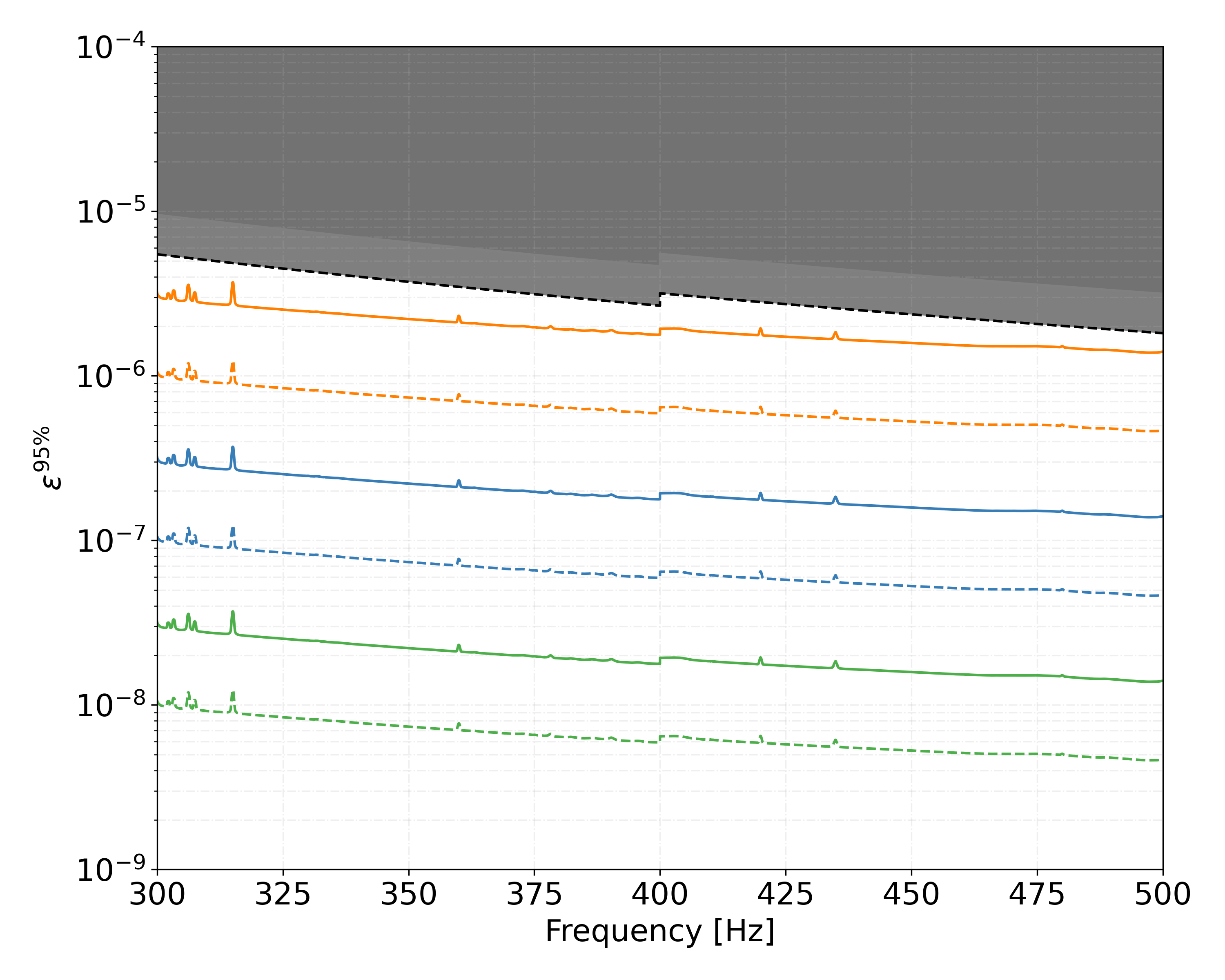}
    \caption{Upper limits on the neutron star ellipticity $\epsilon$ at the $95\%$ confidence level. The three different colors show results for distances of 1\,kpc (upper orange curves), 100\,pc (middle blue curves), and 10\,pc (green bottom curves). The dashed lines use a moment of inertia of $I_{zz} = 3 \times 10^{38}$\,kg\,m$^2$ instead of the canonical $I_{zz} = 10^{38}$\,kg\,m$^2$ value, used by the solid lines. In the upper shaded gray area this search is not sensitive because the high ellipticities would generate a spin-down larger than this search can probe.}
    \label{fig:Reach}
  \end{center}
\end{figure}

The upper limits on $h_0$ can hence be used to calculate upper limits on the asymmetry of the targeted neutron star population by rearranging equation \ref{eq:h0}. These ellipticity upper limits depend on choosing a value for the unknown moment of inertia of the neutron star, which is uncertain by around a factor of three \citep[see section 4B of][]{known_2007}, although more exotic neutron star models such as quark stars or lower-mass neutron stars could have even higher moments of inertia \citep{Horowitz_2010,Owen_2005}. Figure~\ref{fig:Reach} shows these results at three different distances and two values of the moment of inertia. 

These upper limits on the ellipticity are the most restrictive up to date for searches of unknown neutron stars in binary systems. If the true {\it{minimum}} elipticity is $10^{-9}$ as suggested by some studies \citep[such as][]{Woan_Pitkin_Haskell_Jones_Lasky_2018, Gittins_Andersson_2019}, we are about one order of magnitude above that limit for neutron stars at 10\,pc and two orders of magnitude for stars at 100\,pc. For sources at 1\,kpc, with $I_{zz} = 10^{38}$ kg m$^2$ and emitting continuous waves at 500\,Hz, the ellipticity can be constrained at $\epsilon < 1.5 \times 10^{-6}$, while at 10 pc we have $\epsilon < 1.5 \times 10^{-8}$. If we assume $I_{zz} = 3 \times 10^{38}$ kg m$^2$ (as could be due to an equation of state that supports neutron stars with larger radii), the upper limits are more stringent, as shown by the dashed traces. For example, at 100\,pc and 500\,Hz, the upper limit is around a factor of 60 from the claimed minimum at $10^{-9}$. 

An alternative way to present these upper limits is to use the quadrupole moment $Q_{22}$, which is directly related to the $\epsilon I_{zz}$ product :
\begin{equation}
  Q_{22} = \sqrt{ \frac{15}{8 \pi} } \epsilon I_{zz} \rightarrow Q_{22}^{95\%} = \sqrt{ \frac{15}{128 \pi^5} }{c^4\over{G}}{h_0^{95\%} \over {f_0^2}} d 
\end{equation}
By using the quadrupole there is no need to choose a value for the moment of inertia $I_{zz}$. We can compare these upper limits on the quadrupole moment with the results obtained in \citet{maxquadrupolemass}, where the maximum quadrupole moment attainable by a neutron star with a certain mass was calculated. In this way, instead of plotting the absolute upper limit value reached by our search, we can show the ratio by which our search is constraining the maximum value, in a similar way to what is done in targeted continuous gravitational-wave searches where the gravitational-wave amplitude spin-down limit is used \citep{Ashok:2021fnj}.

\begin{figure}
  \begin{center}
    \includegraphics[width=1.0\columnwidth]{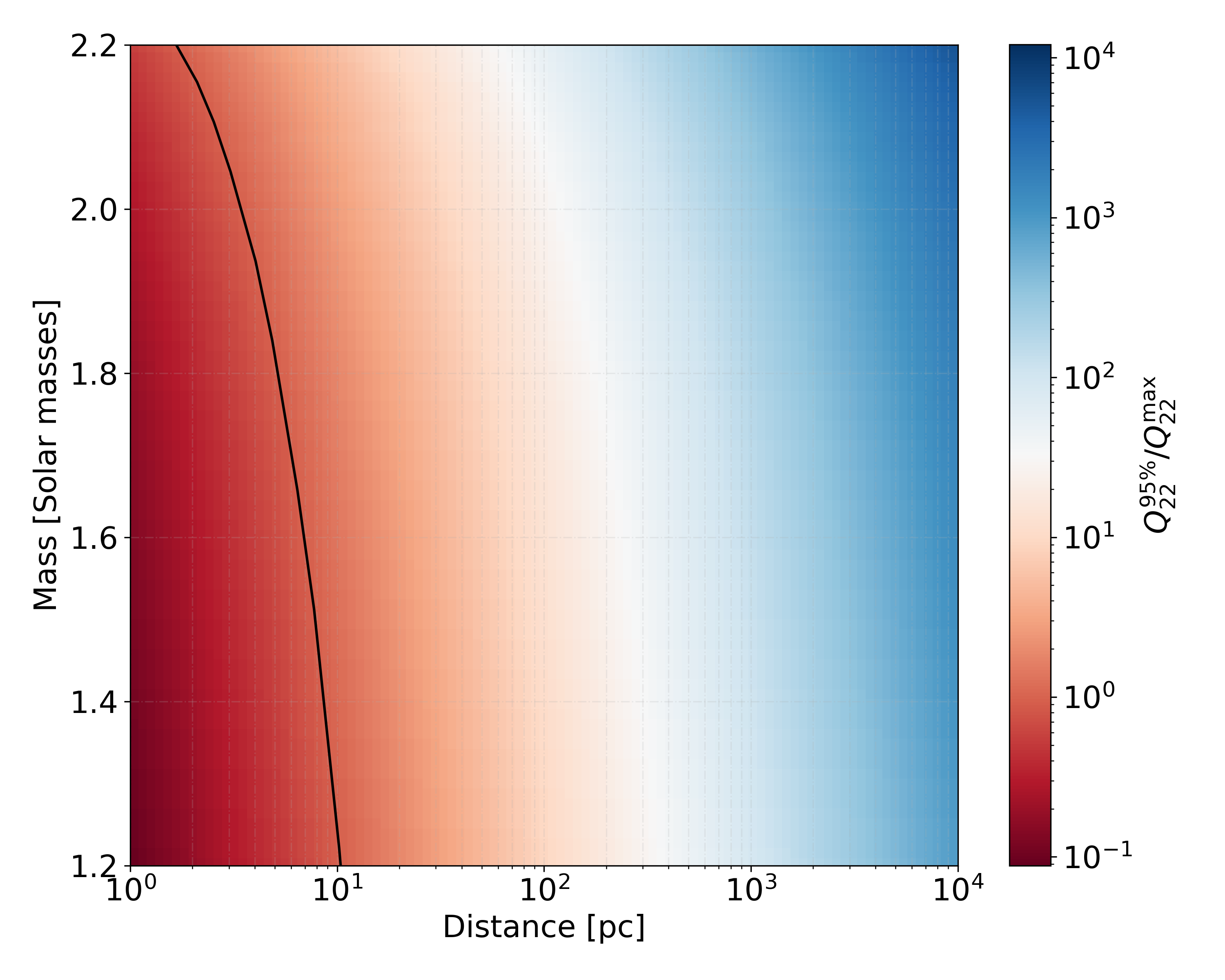}
    \caption{Ratio between our quadrupole moment upper limits $Q_{22}^{95\%}$ and the maximum quadrupole from the upper curve in Figure 1 of \cite{maxquadrupolemass}. The black line marks $Q_{22}^{95\%}/Q_{22}^{\mathrm{max}} = 1$, therefore our upper limits are constraining with respect to the model of \cite{maxquadrupolemass} in the region to the left of this line.}
    \label{fig:Quadrupole}
  \end{center}
\end{figure}

This result can be seen in Figure~\ref{fig:Quadrupole}, where the color indicates the ratio between the calculated upper limits and the maximum quadrupole of \citet{maxquadrupolemass} for a 500\,Hz signal. For a neutron star with 1.2 solar masses, our results exclude maximally strained neutron stars within 10\,pc distance from Earth. Figure~\ref{fig:Quadrupole} also shows that we are more sensitive to neutron stars with lower masses, as previously discussed by \cite{Horowitz_2010}. %\check{If we accept a lower confidence, say 80\%, our constraints extend to 11.5 pc from Earth.}

The considerations above must be taken with a grain a salt, because the maximum quadrupole is highly dependent on the star's unknown formation history, its equation of state, and the breaking strain and shear module of the elastic crust. Furthermore, if the neutron star contains exotic states of matter, the maximum quadrupole moment could be higher \citep{Owen_2005}. For the fiducial value of the moment of inertia of $10^{38}\,{\textrm{kg\,m}}^2$, the quadrupoles of \cite{maxquadrupolemass} correspond to relatively low ellipticities of $\sim 10^{-8}$ and this results in a limited reach in Figure~\ref{fig:Quadrupole}.%Neutron stars with even lower masses would attain higher maximum quadrupole moments \citep{Horowitz_2010}.

\begin{figure}
  \begin{center}
    \includegraphics[width=1.0\columnwidth]{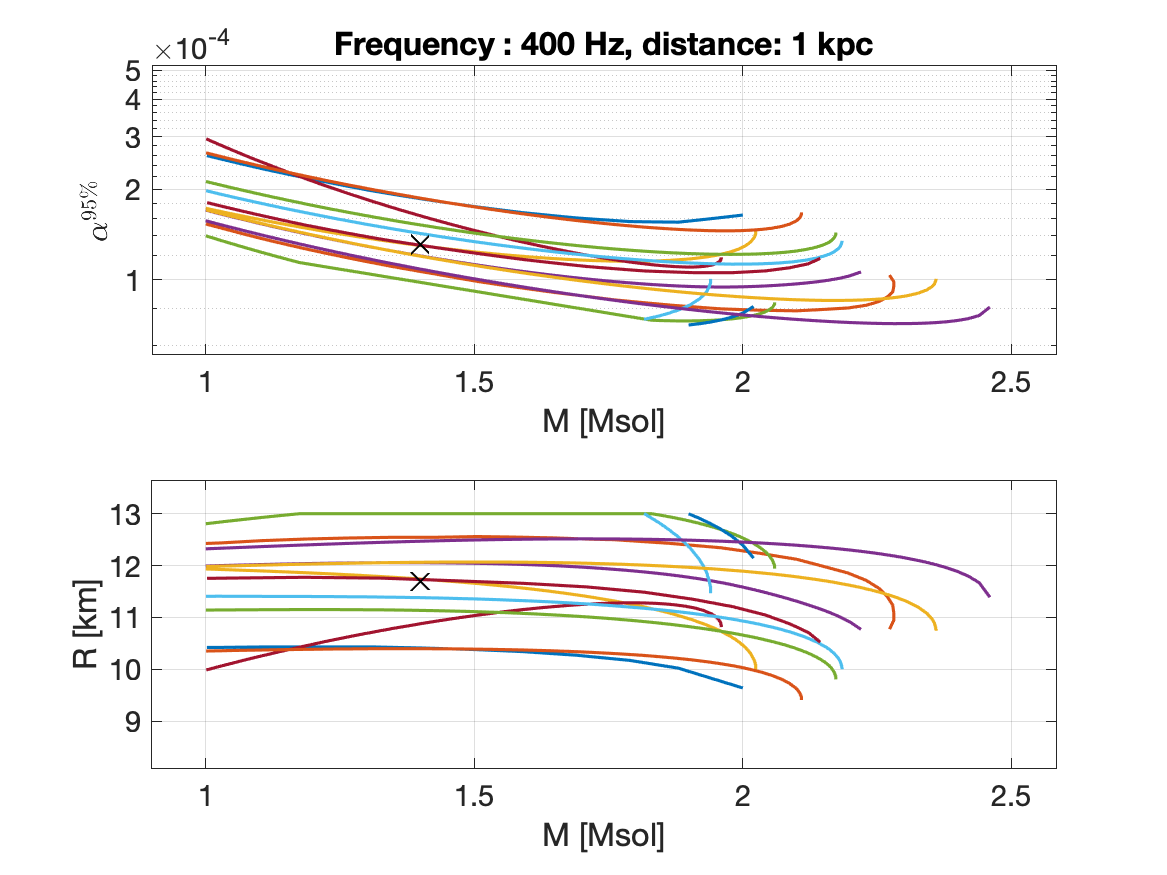}
    \includegraphics[width=1.0\columnwidth]{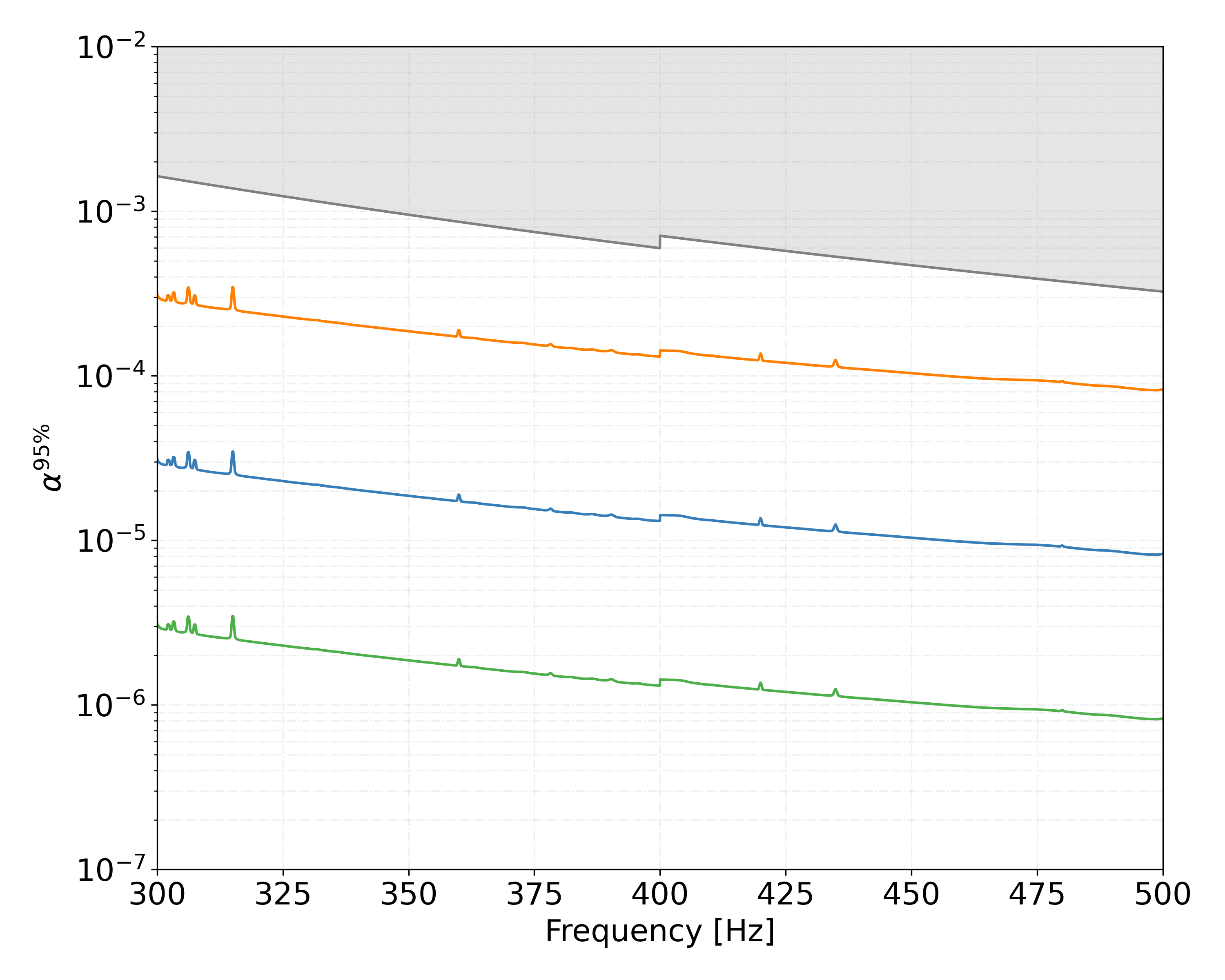}
    \caption{Upper limits $\alpha^{95\%}$ on the r-mode amplitude at the $95\%$ confidence level. {\it {Top plot:}} upper limit as a function of the neutron-star mass, where we assume $f=400$\,Hz and a distance to the source of 1\,kpc. \check{We consider different equations of state to give an idea of the range of variability of the constraints as a function of the unknown mass and radius of the source. {\it {Middle plot}}: mass-radius combinations corresponding to the considered equations of state. The cross indicates the nominal mass and radius values often used of $M=1.4 ~M_\odot$ and $R=11.7$ km (see Eq.~\ref{eq:alpha}).} The equations of state are taken from \cite{Ozel:2016oaf}, {\texttt{http://xtreme.as.arizona.edu/NeutronStars/\-index.php/dense-matter-eos}}, with $M\in [1 , 3] ~M_\odot$ , $R\in [8, 13]$\,km, and only ones with a maximum mass lower than 1.9 $M_\odot$ are included \citep{Antoniadis:2013pzd,Cromartie:2019kug,Hebeler13,Kurkela14}. {\it {Bottom plot:}} upper limit as a function of the signal frequency $f_0$ for sources at 1\,kpc (upper orange curve), 100\,pc (middle blue curve), and 10\,pc (green bottom curve), having assumed $M=1.4 ~M_\odot$ and $R=11.7$ km. The upper shaded gray area shows the region where this search is not sensitive.}
    \label{fig:ReachRModes}
  \end{center}
\end{figure}
If the continuous gravitational waves are emitted by r-modes, apart from a weak dependence on the degree of central condensation of the star, their amplitude is
\begin{equation}
\label{eq:alpha}
\begin{split}
  \alpha = 0.028 \left( \frac{h_{0}}{10^{-24}} \right) \left( \frac{d}{1 \mathrm{~kpc}} \right) 
  \left( \frac{1.4 ~{\mathrm{M}}_\odot} {M} \right) \times \\
  \left( \frac{11.7 ~{\mathrm{km}}_\odot} {R} \right)^3
  \left( \frac{100 \mathrm{~Hz}}{f_0} \right)^{3}.
  \end{split}
\end{equation}
From this Equation we derive upper limits on the r-mode amplitude corresponding to different equations of state, in the range $\alpha^{95\%} \in [3\times 10^{-4} , 7\times10^{-5})]$ at 400\,Hz for sources within a distance of 1\,kpc, see Figure~\ref{fig:ReachRModes}. 
The range of theoretical predictions in the literature on r-mode amplitudes is $\alpha\sim 8\times10^{-7}-10^{-4}$ \citep{Gusakov:2013jwa,gusakov_explaining_2014} or 
  $\alpha\sim10^{-5}$ \citep{bondarescu_spin_2007}.
  %, or $\alpha\sim10^{-3}-10^{-2}$ \citep{bondarescu_spinning_2009}.

\cite{Maccarone:2022bym} have argued that within 1 kpc there should exist a substantial galactic population of fast rotating neutron stars in quiescent LMXBs (qLMXBs). This is intriguing because \cite{Kantor:2021ipr,Gusakov:2013jwa,Chugunov:2014cwu} have suggested the existence of a class of hot fast rotating non-accreting neutron stars that look very much like qLMXBs (HOFNARs) that would support very long-lasting r-modes. Our searches begin to probe such hypotheses.

\section{Conclusions}
\label{sec:conclusions}

%Summary
In this paper we have presented the most sensitive search to date for continuous gravitational waves from unknown neutron stars in binary systems, with gravitational-wave frequencies between 300 and 500\,Hz and orbital periods between 15 and 60 days. We have not detected any astrophysical signal, and we provide the most constraining upper limits in this region of parameter space, improving on existing upper limits by more than an order of magnitude. At spin periods near 4 ms, and distances between 10 and 100\,pc, we are approximately within one order of magnitude of the {\it minimum} ellipticity value of O($10^{-9}$) proposed by \cite{Woan_Pitkin_Haskell_Jones_Lasky_2018}. \check{In this frequency range, a factor of ten improvement in sensitivity is expected with data from the next generation of detectors. Before these come online a factor of a few may be gained with longer coherent segment lengths -- which our new \texttt{BinarySkyHou$\mathcal{F}$} enables.} 

Our r-mode amplitude upper limits are well within the range of saturation values for sources up to 1\,kpc over the entire frequency range, and begin to probe the existence of galactic HOFNARs. 

The amplitude of continuous gravitational waves depends on the square of the frequency, while the noise floor of the detectors increases with a lower power of the frequency, thus making higher frequency searches particularly interesting. On the other hand, for neutron stars in unknown binary systems the resolution in parameter space increases with at least the fifth power of the frequency, making high-frequency searches a real challenge. These results demonstrate that it is now possible to explore the high-frequency range at interesting sensitivity depths. 

\begin{acknowledgments}

The authors thank the scientists, engineers and technicians of LIGO, whose hard work and dedication produced the data that made this search possible.

We thank Benjamin Steltner for the application of the gating method to the analyzed dataset, and Nils Andersson and Fabian Gittins for helpful comments and for the data used to produce Figure~\ref{fig:Quadrupole}. 
This work has utilised the ATLAS cluster computing at MPI for Gravitational Physics Hannover. B.J. Owen's research for this  work was supported by NSF Grant No. PHY-1912625.

This research has made use of data or software obtained from the Gravitational Wave Open Science Center (gw-openscience.org), a service of LIGO Laboratory, the LIGO Scientific Collaboration, the Virgo Collaboration, and KAGRA \citep{gwosc}. LIGO Laboratory and Advanced LIGO are funded by the United States National Science Foundation (NSF) as well as the Science and Technology Facilities Council (STFC) of the United Kingdom, the Max-Planck-Society (MPS), and the State of Niedersachsen/Germany for support of the construction of Advanced LIGO and construction and operation of the GEO600 detector. Additional support for Advanced LIGO was provided by the Australian Research Council. Virgo is funded, through the European Gravitational Observatory (EGO), by the French Centre National de Recherche Scientifique (CNRS), the Italian Istituto Nazionale di Fisica Nucleare (INFN) and the Dutch Nikhef, with contributions by institutions from Belgium, Germany, Greece, Hungary, Ireland, Japan, Monaco, Poland, Portugal, Spain. The construction and operation of KAGRA are funded by Ministry of Education, Culture, Sports, Science and Technology (MEXT), and Japan Society for the Promotion of Science (JSPS), National Research Foundation (NRF) and Ministry of Science and ICT (MSIT) in Korea, Academia Sinica (AS) and the Ministry of Science and Technology (MoST) in Taiwan.

\end{acknowledgments}

%% For this sample we use BibTeX plus aasjournals.bst to generate the
%% the bibliography. The sample631.bib file was populated from ADS. To
%% get the citations to show in the compiled file do the following:
%%
%% pdflatex sample631.tex
%% bibtex sample631
%% pdflatex sample631.tex
%% pdflatex sample631.tex
\bibliography{bib}{}
\bibliographystyle{aasjournal}

\end{document}